# COMPETITION OF SELF-ORGANIZED ROTATING SPIRAL AUTOWAVES IN A NONEQUILIBRIUM DISSIPATIVE SYSTEM OF THREE-LEVEL PHASER

D.N. Makovetskii
Usikov Institute for Radiophysics and Electronics of the National Academy of Sciences of Ukraine
12 Ac. Proskura St, Kharkov, 61085, Ukraine
Phone: +38-057-7203452, Fax: +38-057-3152105, E-mail: makov@ire.kharkov.ua

Understanding of the nature of birth, evolution, competition and dynamical stabilization (or destruction) of spatio-temporal dissipative structures (DS) is of great importance to various areas of modern physics, as well as to significant branches of chemistry, biology etc. The phenomenon of self-organization of DS in nonequilibrium systems is nowadays a subject of intensive experimental investigation and computer-oriented modeling.

Rotating spiral waves (RSW) represent a universal class of robust DS observed not only in diverse fields of the fundamental science, but in several types of man-made active media (or arrays of units) used e.g. in quantum electronics [1] or nanoelectronics [2]. All these RSW have the certain topological charges similar to charges of singular optics vortices. But, in contrast to *conservative* singular optics vortices, our *dissipative* RSW are self-sustained autostructures, namely autowaves. They possess a responsibility of non-decaying propagation in active media (maintained by the balance of an appropriate pumping and corresponding dissipation of energy).

Recently, a possibility of self-organization of DS, and, especially, of robust RSW in autonomous microwave phonon lasers (phasers) [3] has been shown by means of computer experiments [4,5]. These experiments were primary inspired by the theoretical conception of excitability (CE), which is applicable not only to chemical and biological active media [6,7], but to class-B lasers [1] and paramagnetic phasers [3,4] as well.

In class-B lasers, the relaxation of excitation of active centers (AC) is much slower than resonator's active mode(s) decay. Generally, the class-B dissipative systems are characterized by the inequality $\tau_a^{(S)} >> \tau_f^{(L)}$. Here $\tau_a^{(S)} = \min\{\tau_a\}$; $\tau_f^{(L)} = \max\{\tau_f\}$; $\{\tau_a\}$ is a set of the AC longitudinal relaxation times, and $\{\tau_f\}$ is a set of lifetimes of the field (quasi)particles, which are resonantly interacting with AC. The same inequality takes place for phasers [3,4], so it is possible to use CE for modeling DS and, in particular, RSW in phasers [4,5].

In this report, we review and discuss recent results of numerical investigations of RSW dynamics in a nonequilibrium dissipative medium of three-level paramagnetic phaser under assumption $\tau_f^{(L)}/\tau_a^{(S)} \to 0$. Most of the discussed below numerical experiments were fulfilled by using the cross-platform software package TLM (Three-Level Model of excitable system) [5,8,9] and it's previous variant TLCA (Three-Level Cellular Automaton) [10]. The main attention was devoted to phenomena caused by increasing/decreasing competition of self-organized RSW, their coexistence with nonspiral DS, dynamical stabilization/decay of RSW or full collapse of excitations in wide range of quantity of active particles $D$ and control parameters sets $\{C_m\}$.

The TLM package is a Java-based implementation [8,9] of the paramagnetic-oriented modification [4] of the well-known Zykov-Mikhailov (ZM) model [6,7]. The underlying algorithm of TLM (ATLM) [5,8,9] is based on binary/integer/rational numbers and Boolean/arithmetic operations only, so it allows direct mapping of the model onto an architecture of a standard digital computer. The ATLM belongs to the polynomial-time (*P*-time) class of algorithmic complexity; moreover, it is a *Linear*-time algorithm for the case of predefined quantity of iterations $N$ of a modeling session, so it provides possibility of modeling of long evolution of large systems ($ND \approx 10^{12}$ at usual personal computers and $ND \approx 10^{13} - 10^{15}$ at mid-power machines) during a workday. Moreover, the ATLM is intrinsically fine-grained parallel algorithm, based on cellular automata, which resolve in a natural way the contradiction between efficiency and parallelization in computational mathematics. This property of cellular automata gives possibility of further increasing of $ND$ [9], e.g. by using clusters of computers or by specialized co-processors with fine-grained hardware architecture and built-in firmware.

The ATLM [5,8,9] and the corresponding TLM software package [8,9] explicitly include a mechanism of spatially-dependent local inhibition of AC excitations in diluted paramagnetic excitable system [4,5]. The original ZM model [7] takes into account only the usual spatially-independent mechanism of inhibition (which is an analogue of spin-lattice relaxation in paramagnetic systems), so the chemical-oriented ZM model of excitable systems [6,7] disregards any cross-relaxation phenomena which are typical for paramagnetic excitable systems. It must be pointed out that not only ZM model, but almost all models of Wiener-Rosenblueth type [7] ignore spatial dependence of inhibition, restricting itself by spatial dependence of local activation only. This approach is usually correct for chemical systems (e.g. of Belousov-Zhabotinsky type), but it is inadmissible for paramagnetic excitable systems. In works [4,9], it was shown that taking into account both local activation and local inhibition of excitation is equal to generalization of the model with the 1-channel (1C) diffusion to the 2-channel (2C) one.

In computer experiments fulfilled by S.D. Makovetskiy and the author, there were observed several new scenarios of self-organization, competition and dynamical stabilization of RSW under conditions of cross-relaxation between AC [4]. Subsequent careful investigations carried out by S.D. Makovetskiy [5,8,9] on excitable media with 2C-diffusion showed that interaction of RSW with absorbing boundaries possess peculiarities, which was unknown for such media with 1C-diffusion. The most interest results obtained in [5,8,9] – revealing of regeneration and replication of RSW in excitable systems which model active media of paramagnetic phasers.

Regeneration of RSW in a surface layer of bounded excitable medium looks as "reflection" of spiral autowave from a boundary. This, of course, is not a trivial reflection because of non-reflective (fully absorbing) nature of the boundary. Moreover, in contrary to ordinary waves, autowaves generally can not be reflected in usual way (even by perfect mirror). As a matter of fact, nonlinear "reflections" observed in computer experiments [5,8,9] were identified by S.D. Makovetskiy as complex *nonlinear transformations* of RSW leading to inversion of direction of drift of a revived and self-reconstructed RSW. In some cases, non only this direction of RSW core motion, but the sign of the RSW's topological charge $Q_T$ was inverted too [5,8,9]. This last result correlates with observations of self-organized inversion of $\text{sgn}\, Q_T$ in preceding computer experiments [4,10].

Replications of RSW [5,8,9] (increasing of quantity of spirals after their nonlinear transformation in the surface layer) is a higher-order phenomenon of the same nature, as regeneration of RSW. Regenerations and multiple replications of spiral autowaves clarify phenomenon of transient Zeno-like spatio-temporal chaos (TZSChaos) which appeared to be typical [5,9] in investigated excitable media with 2C-diffusion (i.e. with competing local activation and local inhibition of AC excitations). In spite of obvious regularity of all the possible attractors in the TLM model (as in any spatially-bounded cellular automata model), the time of transition process may reach giant values by *almost* precise repetitions of spatial configuration (Zeno-phenomenon). These repetitions are the consequence of nearly perfect balance between quantity of RSW moving to and from boundaries.

It is important that all the computer experiments [4,5,8-10] were performed in *isotropic and homogeneous* (both in the J. von Neumann sense) media. So the mechanism of self-reconstruction of spiral autowaves in excitable media accompanied by nonlinear "reflection" of RSW and inversion of $\text{sgn}\, Q_T$ is quite different from superficially similar phenomena of non-autowave optical vorticity in strongly inhomogeneous media [11,12].

Results of computer experiments on nonstationary phenomena in excitable media emulating paramagnetic systems shed light upon an old but yet non-resolved puzzle of spin-lattice relaxation (SLR) formulated by J.H. Van Vleck in his seminal report at 1st Symposium on Quantum Electronics [13] and widely discussed in many subsequent papers concerning particularly the problem of the nature of bottlenecked SLR in dilute paramagnetics at low temperatures. In most of these papers, the concept of macroscopic (averaged, thermodynamical etc.) description of bottleneck phenomena in SLR was used. A standard model of bottleneck in SLR is the phonon bottleneck: SLR is slowed down by lack of resonant phonons which pass energy from AC to the (thermal) reservoir. At the same time, experiments showed inadequacy of this model. In other words, a description of multi-particle paramagnetic system by small set of macroscopic parameters is unsatisfactory.

Taking in account a possibility of spontaneous formation of DS in paramagnetic media, one can suppose that a bottleneck may originate at the level of particle-particle interaction rather than at the level of particle-reservoir interaction. Computer experiments [4,5,8-10] demonstrate a possible scenario of bottlenecked evolution for an ensemble of interacting three-level AC. Note that this slowing down in SLR is *non-critical* and there is no necessity of fine tuning of the system. On the other hand, such the scenario of bottlenecked SLR is entirely self-organized one, so no external driving (as in SOC dynamics [14]) is required to support TZSChaos.

Spatio-temporal dynamics of RSW is sensitive not only to presence of absorbing boundaries in our isotropic and homogeneous nonequilibrium dissipative system, but to the relationship between a linear dimension $d_L$ of the active medium and the Wiener wavelength $\lambda_W$, which is equal to the distance between neighboring coils of a RSW. For moderate values $d_L/\lambda_W$ (of order 10), there exist robust spiral autowaves with $|Q_T| \geq 1$. But for large values $d_L/\lambda_W \gg 10$, only the spirals with $|Q_T| = 1$ usually retain their robustness, and multispiral DS consist of spirals with this minimal topological charge (note that DS with $Q_T = 0$ are not spirals, they are called pacemakers or target patterns). This is yet another important difference between homogeneous and inhomogeneous (perforated) active media, because in media with holes may exist robust spirals with $|Q_T| \leq p_H/\lambda_W$, where $p_H$ is the perimeter of a hole.

Complex early-stages DS in excitable media with 2C-diffusion gradually evolve to more or less pure vortex states (formed by competing RSW), which may be treated as aggregate states, e.g. of such the types:
- Wigner crystal state (WCS) with frozen positions of all the RSW cores, similarly to a Wigner aperiodic crystal;

- Glass state (GS) with huge quantity of attractors and superslow ageing of the spatially-temporal structure;
- Evaporating liquid state (ELS) with short range ordering, quick drift of RSW cores and decreasing quantity of RSW $K_{\text{RSW}}$, down to $K_{\text{RSW}} = 1$, and, typically, to $K_{\text{RSW}} = 0$ with subsequent entire collapse of excitations;
- Nonevaporating liquid state (NLS), which is similar to ELS, except of $K_{\text{RSW}}(t) \approx \text{const}$, $K_{\text{RSW}}(t \to \infty) \gg 1$.
- Dense plasma state (DPS) with strongly distorted (from Archimedean RSW) "ionized" autowaves, which produce complex, unstable, quickly oscillating irregular patterns of vortices of both RSW and non-RSW type.

Fully developed WCS are characterized by dynamically stable coexistence of RSW with $\text{sgn}\, Q_T = \pm 1$ (diminished competition between RSW). Most of GS at short time intervals looks as frozen-RSW-cores structures (similarly to WCS), but at large time-scales patterns of RSW evolve in a very special manner. Evolution of ELS usually is ended by collapse of excitations over all the active medium, but intermediate stages of evaporating vortex liquid are of great interest because they include processes of spontaneous inversion of $\text{sgn}\, Q_{\Sigma T}$, where $Q_{\Sigma T}$ is the sum of $Q_T$ over all the active medium. Evolution of NLS proceeds under conditions of the TZSChaos (with almost conserved $K_{\text{RSW}}$) and need further investigations due to giant duration of transient stage. Inversion of $\text{sgn}\, Q_T$ in NLS is local and fast, but $Q_{\Sigma T}$ is changed very slowly. And, finally, DPS is the most complex and intriguing state of excitable system, which be investigated in details in subsequent papers.

A special interest represents a group of phenomena of forming and decay of the chimera states (coexisting spatio-temporal structures with incongruous dynamics), which were observed in excitable systems [4]. Chimera state does not belong to any of described above uniform aggregate states. Chimera states are known for nonlocally coupled oscillators [15,16], i.e. systems consisting of elementary units with inherent oscillatory behavior. In contrast to the oscillatory systems, elementary units of excitable systems are monostable relaxational ones. In other words, an isolated excitable AC has the single non-oscillatory stable stationary state, so any rhythmic transitions between the levels of AC are possible only in an ensemble of interacting excitable units.

From this point of view the emergence of chimera states (e.g. the coexistence of periodically and aperiodically oscillating spatial domains [4]) is rather unexpected property of dissipative system containing only relaxational units. But our real radio-physical experiments with microwave phaser also show [3] coexistence of periodically and aperiodically oscillating spatial domains of excitable paramagnetic system $\text{Cr}^{3+} : \text{Al}_2\text{O}_3$. Consequently, both computer modeling and real radio-physical experiments demonstrate presence of chimera states in excitable nonequilibrium dissipative systems, where a special kind of competition of localized DS ensures their antagonistic but persistent coexistence.

This work was reported at 6-th International Kharkov Symposium "Physics and Engineering of Microwaves, Millimeter and Submillimeter Waves" – see Proc. MSMW'2007 (June 25-30, 2007), Kharkov, 2007, Vol.2, Report VI-16.

An extended version of this work, which includes a detailed description of the computational model see e-Preprint arXiv:0805.1319v1 by S.D. Makovetskiy & D.N. Makovetskii [17] (published May 9, 2008).